\documentclass{aa}
\pdfoutput=1
\usepackage{graphicx}
\usepackage{txfonts}
\usepackage[Symbol]{upgreek}
\usepackage[T1]{fontenc}

\begin{document}

\title{Fibre-optic delivery of time and frequency to VLBI station}

\author{P. Krehlik\inst{1}, \L. Buczek\inst{1}, J. Ko{\l}odziej\inst{1},
        M. Lipi\'nski\inst{1}, \L. \'Sliwczy\'nski\inst{1},\\
        J. Nawrocki\inst{2}, P. Noga\'s\inst{2},
        A. Marecki\inst{3}, E. Pazderski\inst{3},\\
        P. Ablewski\inst{4}, M. Bober\inst{4}, R. Ciury{\l}o\inst{4},
        A. Cygan\inst{4}, D. Lisak\inst{4}, P. Mas{\l}owski\inst{4},
        P. Morzy\'nski\inst{4}, M. Zawada\inst{4},\\
        R. M. Campbell\inst{5},
        J. Pieczerak\inst{6},
        A. Binczewski\inst{7}, K. Turza\inst{7}
}

\authorrunning{P. Krehlik et al.}

\institute{AGH University of Science and Technology,
           Mickiewicza 30, 30-059 Krak\'ow, Poland\\
           \email{krehlik@agh.edu.pl}
           \and
           Astrogeodynamic Observatory (AOS), Borowiec,
           Drapa{\l}ka 4, 62-035 Borowiec, Poland
           \and
           Centre for Astronomy, Faculty of Physics,
           Astronomy and Informatics, Nicolaus Copernicus University,\\
           Grudzi\k{a}dzka 5, 87-100 Toru\'n, Poland
           \and
           Institute of Physics, Faculty of Physics,
           Astronomy and Informatics, Nicolaus Copernicus University,\\
           Grudzi\k{a}dzka 5, 87-100 Toru\'n, Poland
           \and
           Joint Institute for VLBI ERIC,
           Postbus 2, 7990 AA Dwingeloo, The Netherlands
           \and
           Orange Polska S.A., Obrze\.zna 7, 02-691 Warsaw, Poland
           \and
           Pozna\'n Supercomputing and Networking Center,
           Jana Paw{\l}a II 10, 61-139 Pozna\'n, Poland
}

\date{Received 14 February 2017 / Accepted 17 March 2017}

\abstract{The quality of Very Long Baseline Interferometry (VLBI) radio 
observations predominantly relies on precise and ultra-stable time and 
frequency (T\&F) standards, usually hydrogen masers (HM), maintained locally 
at each VLBI station. Here, we present an operational solution in which the 
VLBI observations are routinely carried out without use of a local HM, but 
using remote synchronization via a stabilized, long-distance fibre-optic 
link. The T\&F reference signals, traceable to international atomic 
timescale (TAI), are delivered to the VLBI station from a dedicated 
timekeeping laboratory. Moreover, we describe a proof-of-concept experiment 
where the VLBI station is synchronized to a remote strontium optical lattice
clock during the observation.
}


\maketitle

\section{Introduction}

There is one crucial technical difference between Very Long Baseline 
Interferometry (VLBI) and the interferometric techniques used in 
radio astronomy in case of small (tens of kilometres) or medium-sized 
(hundreds of kilometres) baselines. This difference is that in VLBI the electromagnetic 
signals received by the individual radio telescopes are not correlated 
during the observation itself but ``off-line'', that is, they are recorded,
then the media are shipped to and played back at the correlator to finally 
acquire the values of fringe visibility function. To preserve the mutual 
coherence of the signals recorded at all stations, the receivers and the 
data formatters have to be synchronized to ultra-stable time and frequency 
(T\&F) standards. Each station has its own T\&F standard. The common practice 
is to use hydrogen masers (HM) because they are affordable and their 
short-term stability appears sufficient for this purpose.

With the advent of very fast fibre-optic links, the above paradigm of VLBI 
has changed dramatically. In brief, owing to the availability of high-speed 
links, the VLBI resembles the short-base (``local'') radio interferometry
more than ever before. The breakthrough is implementation of the so-called 
\hbox{e-VLBI}, where the data correlation is performed in real time, meaning
without recording and shipping of the media. But even in \hbox{e-VLBI}, each of
the participating stations still relies on its own T\&F standard to which it is 
synchronized. This, however, may change too. New opportunities arose 
because of the development of novel techniques of high-precision fibre-optic 
distribution of T\&F signals over distances of hundreds of kilometres. The 
advantages of this approach can be summarized as follows: (1) stations may 
be synchronized by remote T\&F standards so they no longer need to install 
and maintain their own, (2) many stations may receive synchronization signal 
traceable to a common international atomic timescale (TAI), and (3) some 
stations may get access to new generation optical atomic clocks, 
outperforming traditional HMs in terms of both stability and absolute 
accuracy. As optical atomic clocks are rapidly becoming the most stable 
clocks and have already significantly surpassed their microwave counterparts 
\citep{Hinkley2013,Targat2013,Bloom2014,Falke2014,Ushijima2015}, the last
point emerges as particularly applicable to an actively developed
millimetre-wavelength VLBI, already limited by the stability of HMs for required
integration times \citep{Rioja2012,Rogers1981,Thompson1986}.

\begin{figure*}
\includegraphics[width=\linewidth]{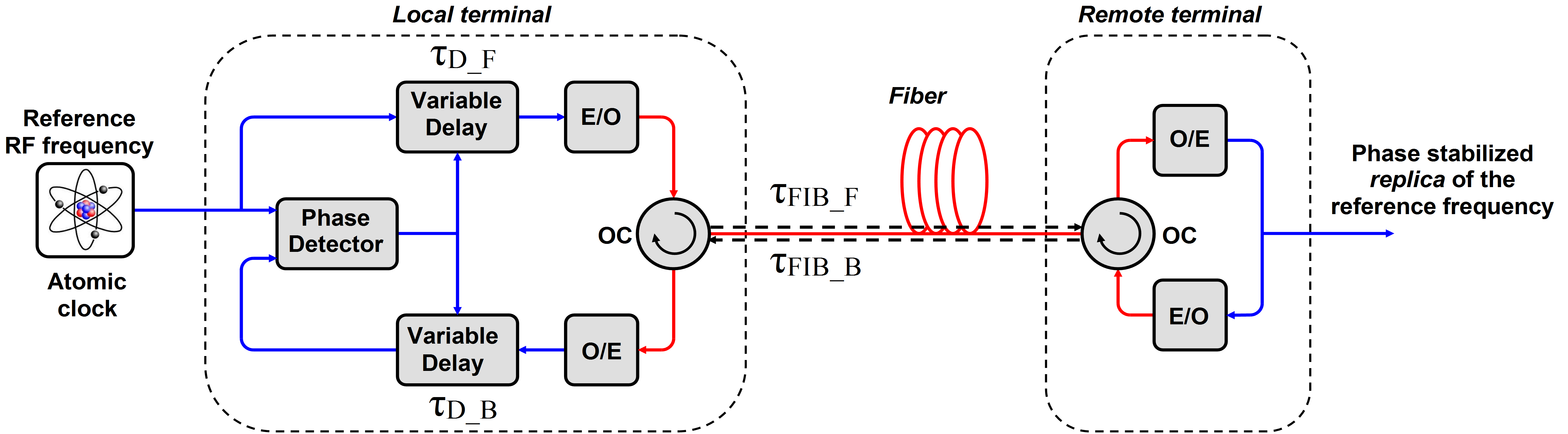}
\caption{Simplified block diagram of the ELSTAB system; E/O stands for an 
intensity-modulated laser, O/E -- an optical-to-electrical receiver, OC --
an optical circulator.}
\end{figure*}

\begin{figure*}
\includegraphics[width=0.45\linewidth]{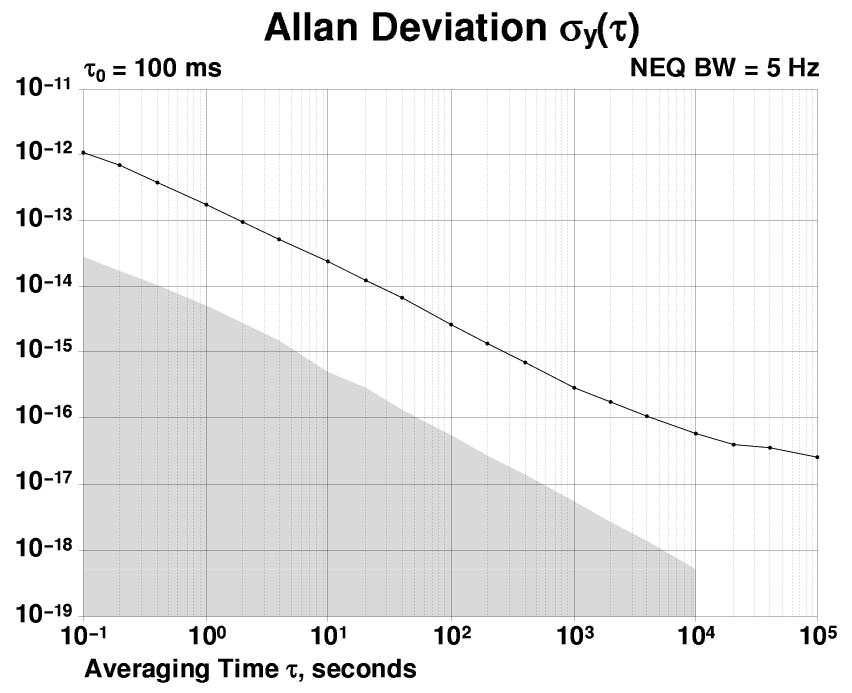}
\includegraphics[width=0.45\linewidth]{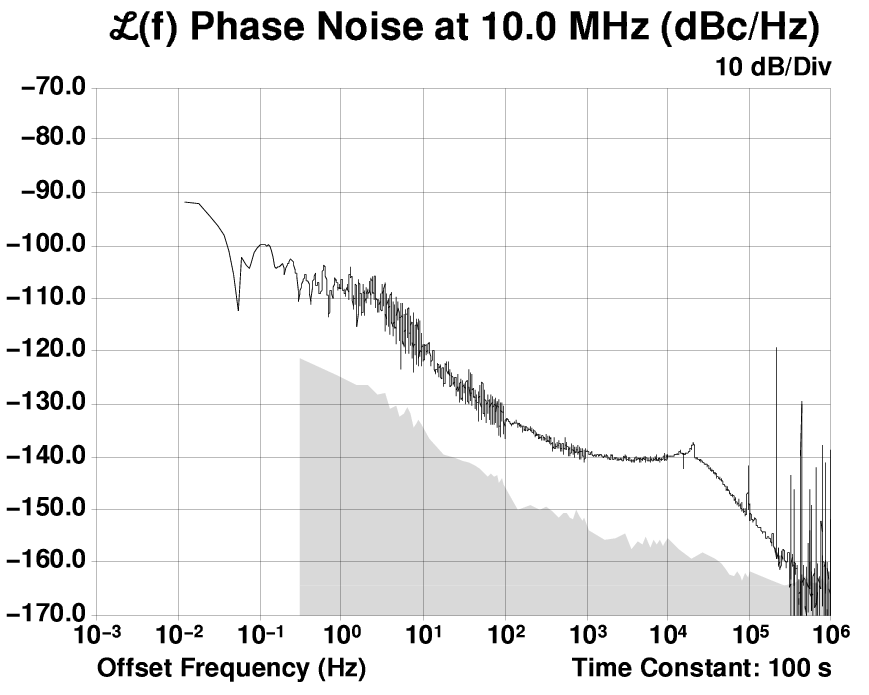}
\caption{Allan deviation (left) and phase noise (right) introduced by ELSTAB.}
\end{figure*}

Here, we present the basics of fibre-optic systems dedicated for 
distribution of T\&F reference signals (Sect.\,2). Next, we describe the 
solution implemented in Poland where a fully operational delivery of 
UTC-traceable T\&F signals from the Time Laboratory of the Astrogeodynamical 
Observatory in Borowiec near Pozna\'n (AOS) to the VLBI station in Piwnice 
near Toru\'n operated by the Centre for Astronomy of Nicolaus Copernicus 
University in Toru\'n (hereafter: Toru\'n VLBI station) has been established 
over a 345-km-long fibre link (Sect.\,3). In Sect.\,4, we present a 
proof-of-concept experiment where the Toru\'n VLBI station was synchronized 
with a remote strontium optical lattice clock developed at National 
Laboratory for Atomic, Molecular and Optical Physics (KL\,FAMO). Our results 
are summarized in Sect.\,5.

\section{Fibre optic distribution of frequency and time}

During the last decade, optical fibre has been recognized as a very powerful 
medium for long-distance distribution of the highest quality T\&F signals, 
generated in metrological institutions dedicated to developing and 
maintaining atomic clocks and generating atomic timescales. Many such links, 
at both experimental and operational levels, have been recently established 
\citep{Lopez2010,Lopez2012,Droste2013,Sliwczynski2013,Calonico2014,Dierikx2016}.
There are, however, specific issues related to 
vulnerability of the fibres to environmental factors that should be 
addressed to ensure high stability of delivered synchronization signals. 
The most pronounced effect is the thermal instability of the propagation 
delay that affects both phase and frequency of the carrier signal and also 
the temporal positions of received time-stamps. The reason for this effect is 
mainly the thermal dependence of the refractive index of silica glass and – 
to a smaller extent – variation of the physical length of the fibre. For 
various types of telecommunication-grade fibres the related coefficient of 
the delay fluctuations is close to 40\,ps\,K$^{-1}$\,km$^{-1}$ \citep{Sliwczynski2010}. Other
important environmental factors are variable tensions and vibration affecting
the cable that also modulate the propagation delay. The power spectral density
of this ``acoustic noise'' is usually located in the range between 10 and
40\,Hz and is strongly dependent on the fibre location: the noise is much more
pronounced in urban areas and especially in fibres deposited along railways.
A very specific case is also aerial-suspended fibres, directly exposed to
diurnal temperature fluctuations and solar heating, winds, and vibrations
generated by high-voltage power lines \citep{Sliwczynski2016}.

The common method used to overcome these environmental limitations is to arrange 
a feedback system in which the signal is sent to the remote (user) side, 
where it is redirected backward to cancel the delay fluctuations in a closed 
loop topology. The cancellation of the delay fluctuations may be performed 
directly in the optical or electrical domain \citep{Ma1994,Lopez2010,Droste2013,Sliwczynski2013},
or realized at the software (protocol) level \citep{Dierikx2016}.

\begin{figure*}
\includegraphics[width=\linewidth]{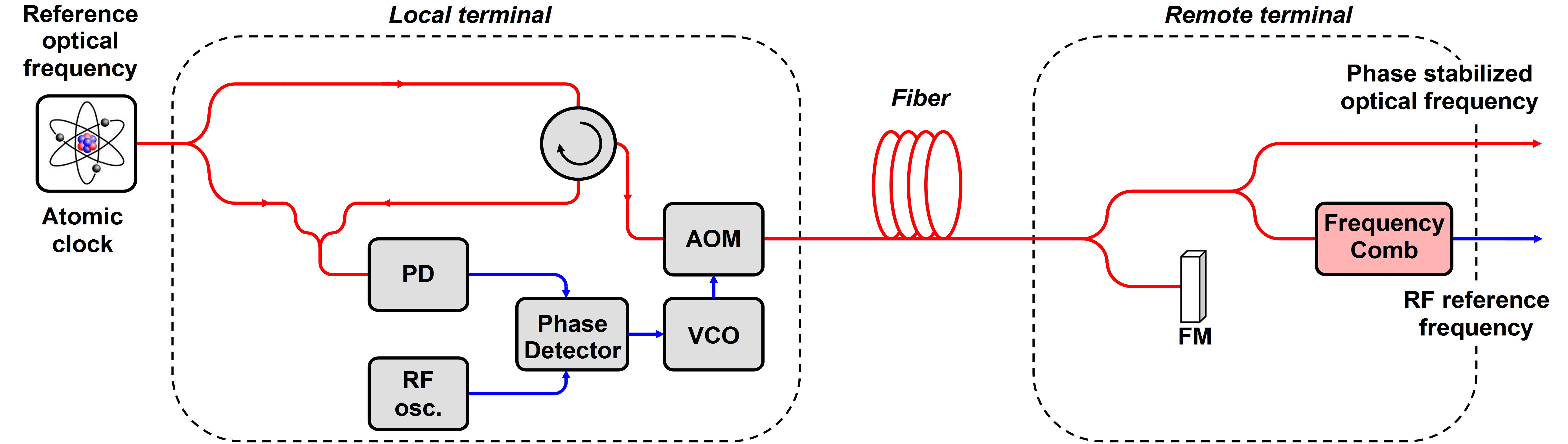}
\caption{Simplified block diagram of the stabilized distribution of the 
optical carrier frequency; AOM stands for an acousto-optic modulator, VCO – 
a voltage controlled oscillator, PD -- a photo-diode, and FM -- a Faraday 
mirror.}
\end{figure*}

In our solution called ELSTAB (see Fig.\,1), the fluctuations of propagation 
delay are compensated electronically by means of the two precisely matched 
variable delay lines and the phase detector that controls the delay lines to 
keep the round trip phase relation (and thus also the round trip delay) 
constant, unaffected by the fluctuations of the fibre delay. This may be 
written as: 
\begin{equation}
\uptau_{\rm D\_F}+\uptau_{\rm FIB\_F}+\uptau_{\rm D\_B}+\uptau_{\rm FIB\_B}={\rm const.}, 
\end{equation}
where $\uptau_{\rm D\_F}$ and $\uptau_{\rm D\_B}$ are (variable) delays of the delay 
blocks and $\uptau_{\rm FIB\_F}$, $\uptau_{\rm FIB\_B}$ are the propagation delays 
of the fibre for the forward and backward directions, respectively. Assuming 
that the fibre delay variations are the same in both directions 
\begin{equation}
\Delta\uptau_{\rm FIB\_F}=\Delta\uptau_{\rm FIB\_B}
\end{equation}
and that the tuning characteristic of both delay blocks are identical
\begin{equation}
\Delta\uptau_{\rm D\_F}=\Delta\uptau_{\rm D\_B},
\end{equation}
it may also be noticed that the one-way delay from the 
input to the remote output, determined by the sum: 
$\uptau_{\rm D\_F}+\uptau_{\rm FIB\_F}$, is constant. (For more detailed analysis and 
discussion, see \citet{Sliwczynski2011}.) The typical stability of the 10\,MHz frequency 
transfer is shown in Fig.\,2. The system-induced Allan deviation is about 
$2\!-\!4\times10^{-13}$ for one-second averaging and goes down with a slope 
close to $\uptau^{-1}$ where $\uptau$ is an observation period. The 
introduced phase noise is about $-$105\,dBc/Hz for a 1\,Hz offset and about 
$-$140\,dBc/Hz for a 1\,kHz offset. Additionally, the full ELSTAB system also 
comprises the time distribution facility, achieved by embedding the 
time-stamps (1\,PPS signal), whose offset (delay) with respect to the UTC 
reference may be calibrated with an uncertainty well below 100\,ps \citep{Krehlik2012}.
In the case of long-haul links, dedicated bidirectional optical amplifiers should 
be installed to compensate the fibre attenuation \citep{Sliwczynski2013}.

A fundamentally different approach for frequency distribution is usually 
used when an optical clock is the source of reference signal. In this case, 
the optical frequency generated in the clock system is directly launched 
into the fibre and basically no modulation is applied. As the environmental 
factors that influence the fibre affect the optical carrier frequency observed 
at the output, such a link should also be stabilized. The typical solution 
is presented in Fig.\,3.
The optical carrier is transmitted to the remote terminal, where some part 
of the signal is redirected backward using a Faraday mirror. The frequency 
of the light coming back to the local terminal is compared with the 
reference provided directly from the clock, thus the frequency
fluctuations induced in the fibre may be detected and corrected. The 
corrections are performed with an acousto-optic modulator that shifts the 
phase/frequency of the light going both in forward and backward directions 
\citep{Droste2013,Calonico2014}. Such systems allow full exploitation of the
superior stability of 
the optical clocks, but it should be borne in mind that the ``natural'' 
output of the system is the 190\,THz optical frequency that cannot directly 
be used in the electrical domain. It may be converted to electrical RF 
domain by an optical frequency comb \citep{NS2007}, this, however, seriously 
increases the cost and complexity of the overall installation. Nevertheless, 
first experiments using this technique for characterization of a HM at the 
Medicina VLBI station \citep{Clivati2015} and to synchronize a VLBI session
\citep{Perini2016,Clivati2017} have been performed.

From a practical point of view, the main concern is the compatibility of the 
frequency and time distribution with other services carried out in the 
fibre-optic network. The simplest solution, applicable when spare fibres are 
available in an optical cable, is to use such a fibre solely for frequency 
and time distribution -- see Fig.\,4a. In the case of a long-distance link 
(longer than approximately 100\,km), bidirectional optical amplifiers should 
be installed at spacings of 40--70\,km. This solution, although not effective 
in terms of fibre usage, is preferred by network operators as it 
does not interfere with standard telecommunications services. Sharing the 
fibre used for other services is also possible by means of wavelength 
division multiplexing (WDM), commonly used in optical networks. Typically, 
the wavelength grid is organized with 100\,GHz or 50\,GHz channel spacing and 
one or two channels needed for frequency and time distribution may be 
selected by optical filters (add-drop multiplexers) and processed 
(amplified) independently to other wavelengths dedicated for standard 
services -- see Fig.\,4b. It should be stressed, however, that in long-haul 
optical networks fibres are used in pairs, one fibre for each direction, 
whereas in our case the preferable option is to use the same fibre for both 
directions. Even though the solution depicted in Fig.\,4b was demonstrated to 
be safe for telecommunication traffic and to maintain frequency and time 
distribution stability \citep{Lopez2012}, it is still hardly accepted within the 
community of network operators. The last option in the case of lack of spare 
fibres is to use so-called alien-lambda scenario where the 
synchronization signals are transmitted on dedicated wavelength in two 
parallel fibres, and routed and amplified by a standard telecommunication 
equipment located in the network nodes, similarly as other services -- see 
Fig.\,4c. A serious drawback of this solution is degradation of the 
forward-backward propagation symmetry that was the underlying assumption of 
the stabilized frequency and time distribution; the behaviour of parallel 
fibres located in the same cable can still be correlated, but distinct 
network equipment in the nodes for each direction causes severe stability
degradation \citep{Sliwczynski2015}.

\begin{figure*}
\includegraphics[width=\linewidth]{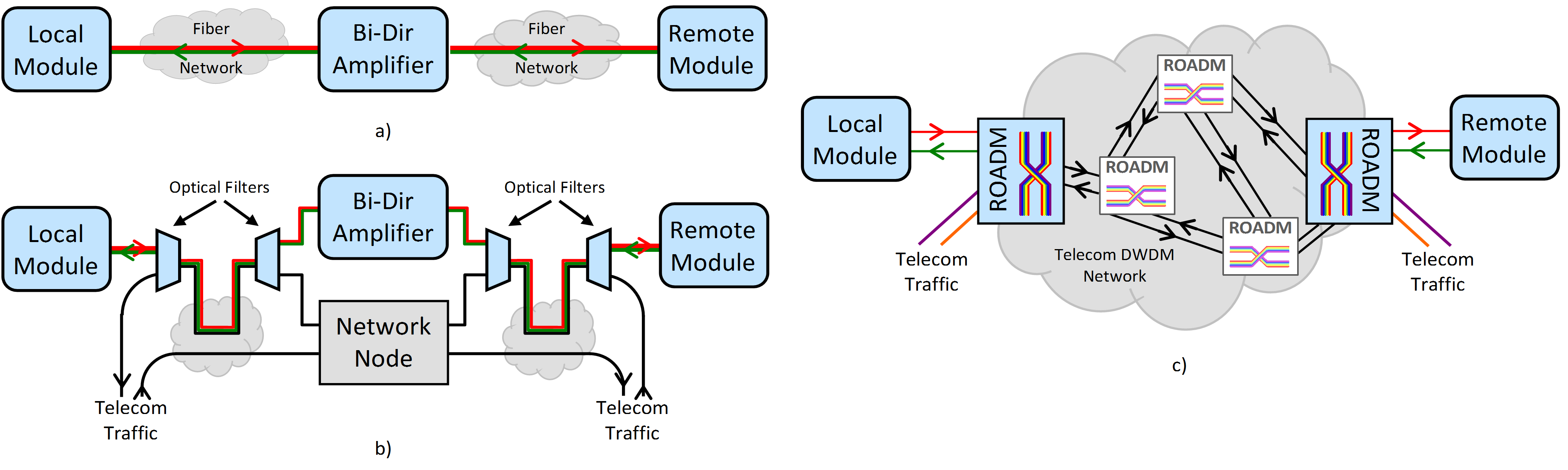}
\caption{Possible strategies for frequency and time distribution exploiting 
fibre optic telecommunication infrastructure: dark-fibre approach (a), 
dark-channel approach (b) and alien-lambda approach (c). ROADM stands for 
re-configurable optical add-drop multiplexer.}
\end{figure*}

\begin{figure}
\includegraphics[width=\linewidth]{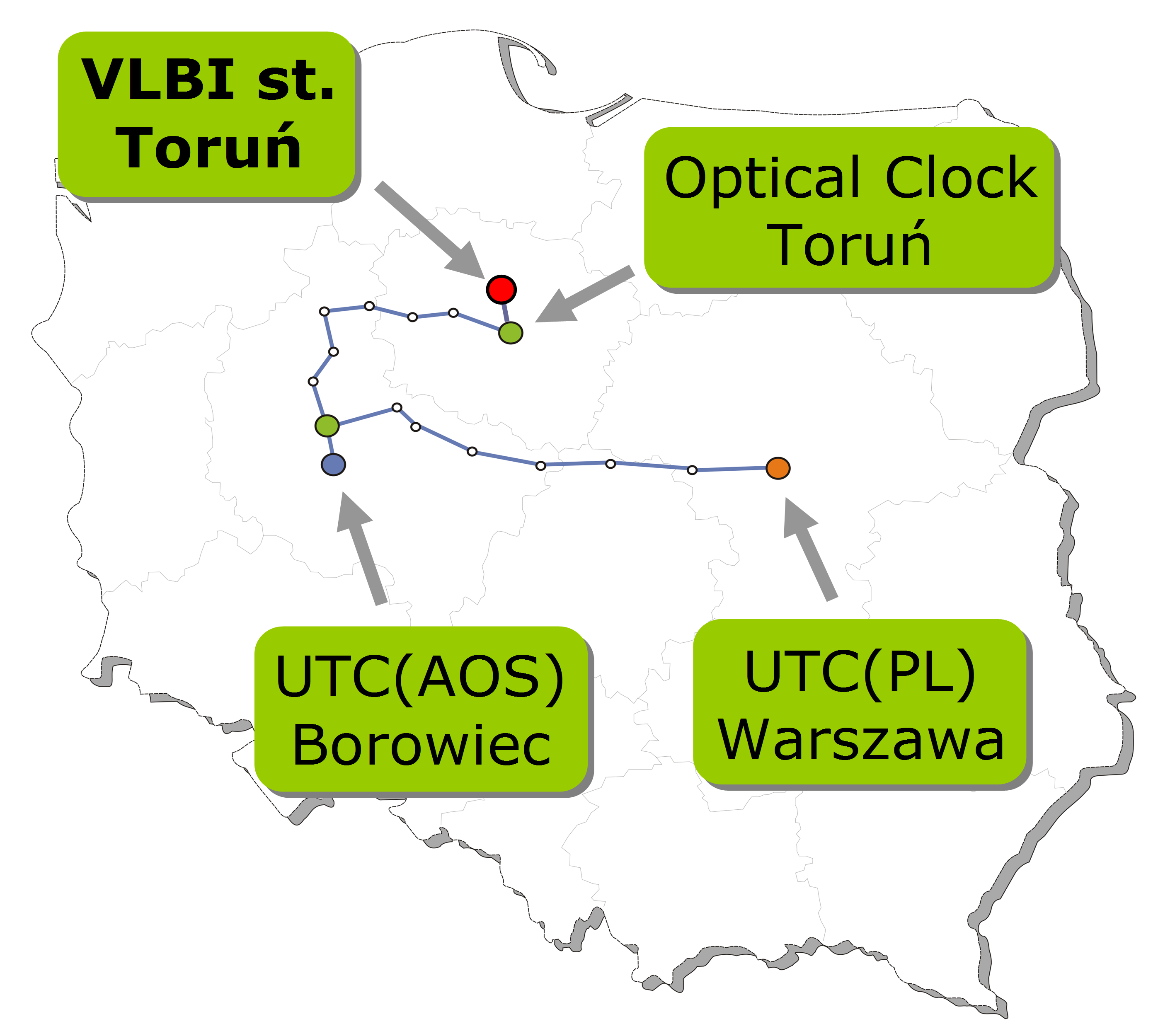}
\caption{VLBI station in Piwnice near Toru\'n connected to the fibre 
optic T\&F distribution network developed under the OPTIME project.}
\end{figure}

\section{UTC traceable, remote synchronization of the Toru\'n VLBI station}

In 2015, the Polish fibre-optic network for T\&F distribution (OPTIME) 
reached Nicolaus Copernicus University in Toru\'n \citep{Buczek2015} and the
Toru\'n VLBI station 
was connected to the backbone with a 15-km-long slave link. This created an 
opportunity to synchronize the Toru\'n VLBI station to one of the Polish UTC 
laboratories, UTC(AOS) or UTC(PL), and alternatively to the strontium optical
lattice clock developed at KL\,FAMO in Toru\'n \citep{Bober2015} -- see Fig.\,5.
The total 
length of the synchronizing path is approximately 350\,km. Physically, the 
T\&F signals are generated at AOS and are based on an HM working as a 
flywheel, but it should be stressed that the HM output is precisely 
corrected to obtain T\&F references -- UTC(AOS) -- coherent with the TAI. 
Therefore, the signals received at the Toru\'n VLBI station are both stable and 
UTC-traceable.

We conducted a dedicated VLBI experiment on 18~December~2015 to serve as a 
proof-of-concept to evaluate the quality of the remote synchronization and 
hence its suitability for operational use in VLBI experiments. Besides the
Toru\'n VLBI station, the participating telescopes included Westerbork (NL), 
Medicina (IT), and Yebes (ES). The observing format comprised four 
single-polarization 8\,MHz sub-bands, together spanning 4974.49\,MHz to 
5006.49\,MHz. The experiment lasted for three hours, continually tracking the 
bright radio quasar 2007+777. The raw data from all telescopes were 
transferred over optical fibre to the Joint Institute for VLBI ERIC (JIVE) 
and the correlation proceeded in real time, that is, using the \hbox{e-VLBI} 
technique.

During the experiment, the quality of the remote synchronization was 
evaluated by dividing the experiment into two domains: first when the T\&F 
signal came from the local HM of the Toru\'n VLBI station and second when 
this was swapped out for the remote HM at AOS provided via fibre. We then 
compared the behaviour of the residual fringe phase in the individual 
baselines/sub-bands in each ``domain''. In addition to being correlated in the 
\hbox{e-VLBI} mode, about 45\,min of the in-streaming data from the telescopes 
for each of the two domains were also recorded at JIVE to enable 
subsequent re-correlations using different coherent integration intervals. 
The correlator model \citep{Gordon2006} included the standard geometric, relativistic,
and propagation terms, plus a first-order polynomial clock model per telescope. 
The a priori rate coefficients of these clock models were set by linear fits 
to a few day’s worth of monitoring the telescopes’ maser outputs against 
GPS. At the beginning of each domain, we adjusted the clock offset and 
rate coefficients to flatten residual delays and rates by correlating 
a few minutes of data. We draw the reader's attention to this process here because
the local HM at the Toru\'n VLBI station had developed a large rate bias (about
$-$19.9\,ps/s at the time of this experiment) that was removed by this 
adjustment procedure. The remote HM showed no significant residual rate with 
respect to UTC.

Because the two domains are intrinsically distinct time ranges, the phases 
on the baselines to Toru\'n cannot simply be subtracted to compare directly 
the two methods of T\&F control at the Toru\'n VLBI station: Earth rotation will 
lead to a changing projection of a baseline with respect to the radio source 
structure on the plane of the sky and unmodelled propagation effects along 
the lines of sight from the individual telescopes will also evolve (the more 
important effect for the time-scales involved here). Fig.\,6 shows an example 
of the behaviour of the residual interferometer phase, plotting one 15\,min 
scan from each domain, using the middle two sub-bands.

\begin{figure*}
\includegraphics[width=\linewidth]{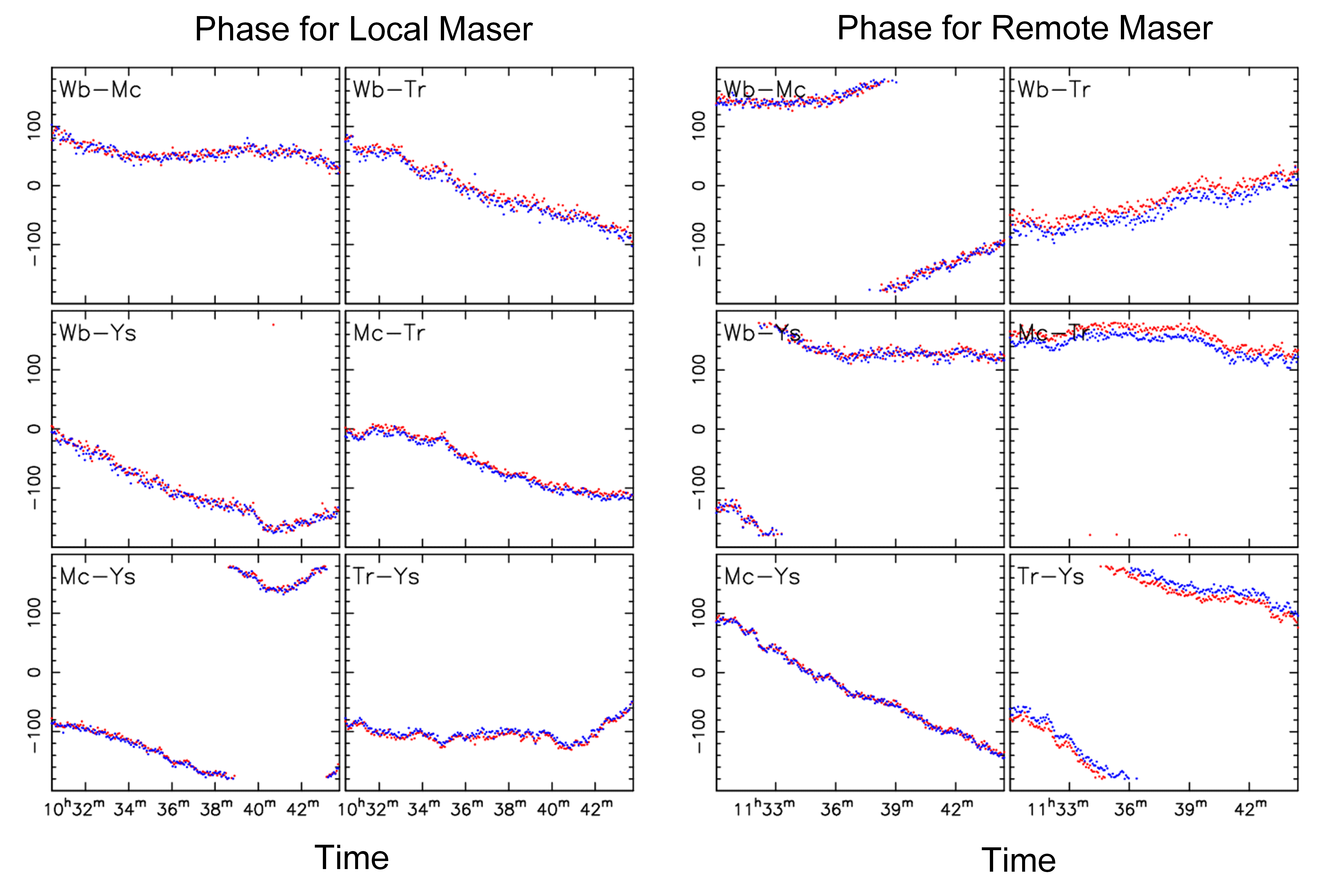}
\caption{Residual interferometer phase for two methods of providing T\&F 
control to the Toru\'n VLBI station during the experiment of 18~December~2015.
The $x$-axis is UTC and the $y$-axis is 
degrees of phase in the middle two sub-bands (colour-coded), vector averaged 
over the middle 80\% of each sub-band. Each four-second integration interval 
is plotted as a separate point. The baselines are arranged such that those 
including Toru\'n are in the right-hand column for each of the T\&F control 
methods.}
\end{figure*}

We considered two phase-difference quantities to characterize the phase 
noise achieved for each method of T\&F control at Toru\'n, using both 2- and 
4\,sec. integration intervals: first, a time-series of the differences of 
consecutive phases, and second, a time series of the difference between a 
phase and the linear interpolation between the phases of the two surrounding 
integration intervals. In each case, the time-series were split into two or 
three subsets, respectively, such that each individual integration 
contributed to only one difference pair or triplet within a given subset. We 
then compared the rms of each subset as a means to characterize the phase 
noise, independent of any remaining unmodelled phase slopes across the full 
time range considered. Table\,1 shows the results for the case of the 
difference-pair statistic with the two subsets labelled as ``odd'' and 
``even'' referring to the parity of the first point in each pair comprising 
the subset. The relative magnitude of these phase-noise statistics across 
the various baselines is in line with the sensitivities of the constituent 
telescopes. The very similar results for the two domains strongly 
suggests that the phase noise is dominated by factors unrelated to the 
specific method of T\&F control at the Toru\'n VLBI station and that the remote HM 
at AOS can provide the operational synchronization for VLBI experiments.

Thus, after the successful completion of this experiment, it was decided to 
permanently switch the synchronization at the Toru\'n VLBI station to the 
remote HM. The motivation for this was that the local HM that had provided T\&F 
control for the Toru\'n VLBI station for 20 years had developed the 
aforementioned worsening drift and its maintenance had become troublesome, 
calling into question its operational reliability. In the end, all regular 
VLBI sessions carried out since the beginning of 2016 have been performed 
with the remote synchronization provided over the fibre from UTC(AOS) 
laboratory.

\section{Experimental synchronization with strontium optical lattice clock}

It has been recently shown that in millimetre and sub-millimetre VLBI the 
losses induced by HM instabilities are comparable to those from high-quality 
tropospheric conditions \citep{Rioja2012} and significant improvement in sensitivity
can be expected after replacing a traditional HM by a more stable frequency 
standard, for example, an optical clock. However, the low-noise down-conversion of 
the optical frequency to RF or microwave domain is essential for exploiting 
the superior stability of the optical clock.

\begin{figure*}
\centering
\includegraphics[width=0.8\linewidth]{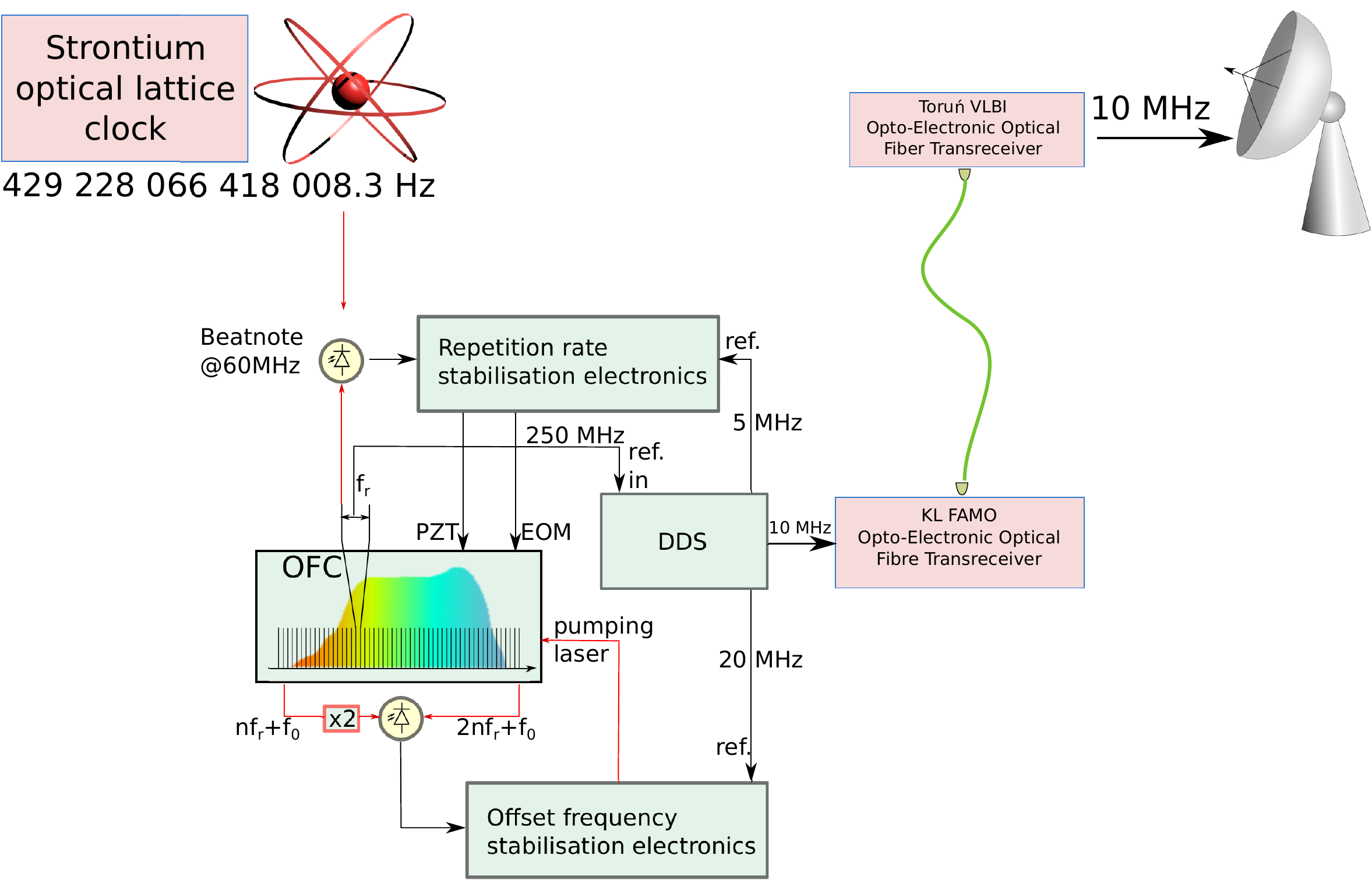}
\caption{Toru\'n VLBI station referenced to a strontium optical lattice 
clock over an optical frequency comb system and a 15-km-long ELSTAB-based 
fibre link. $f_r$ and $f_o$ are the repetition and offset frequencies of
the comb, respectively. OFC stands for an optical frequency comb and DDS
stands for a direct digital synthesizer.}
\end{figure*}

\begin{table}
\caption{Phase noise in picoseconds for different baselines and sub-bands
for local and remote HM.}
\centering
\begin{tabular}{l c c c c c}
\hline
\hline
Baseline & Sub-     & \multicolumn{2}{c}{Local HM}
                    & \multicolumn{2}{c}{Remote HM}\\
         & band     & noise & noise & noise & noise\\
         & and      & rms & rms & rms & rms \\
         & polarity & odd & even & odd & even \\
\hline
Wb-Tr & 0RR & 4.928 & 4.331 & 5.045 & 4.949\\
Wb-Tr & 1RR & 5.062 & 5.130 & 5.152 & 5.186\\
Wb-Tr & 2RR & 5.318 & 5.048 & 5.015 & 5.950\\
Wb-Tr & 3RR & 4.169 & 5.032 & 5.475 & 4.518\\
\hline
Mc-Tr & 0RR & 2.643 & 2.721 & 2.687 & 2.879\\
Mc-Tr & 1RR & 2.929 & 2.605 & 3.151 & 3.230\\
Mc-Tr & 2RR & 2.840 & 2.488 & 2.993 & 3.041\\
Mc-Tr & 3RR & 3.004 & 2.855 & 2.764 & 3.218\\
\hline
Tr-Ys & 0RR & 2.799 & 2.828 & 2.807 & 2.593\\
Tr-Ys & 1RR & 2.825 & 2.234 & 2.610 & 2.507\\
Tr-Ys & 2RR & 2.792 & 2.884 & 2.928 & 3.074\\
Tr-Ys & 3RR & 2.906 & 2.723 & 2.841 & 2.769\\
\hline
Wb-Mc & 0RR & 4.510 & 4.436 & 4.000 & 4.378\\
Wb-Mc & 1RR & 4.719 & 4.486 & 4.144 & 3.929\\
Wb-Mc & 2RR & 4.443 & 4.986 & 3.912 & 4.195\\
Wb-Mc & 3RR & 5.203 & 5.022 & 4.572 & 4.912\\
\hline
Wb-Ys & 0RR & 4.829 & 4.780 & 4.749 & 4.622\\
Wb-Ys & 1RR & 4.403 & 4.608 & 4.781 & 4.454\\
Wb-Ys & 2RR & 4.238 & 3.991 & 4.581 & 4.612\\
Wb-Ys & 3RR & 4.238 & 3.982 & 4.543 & 4.742\\
\hline
Mc-Ys & 0RR & 2.400 & 2.424 & 2.828 & 2.355\\
Mc-Ys & 1RR & 2.234 & 2.286 & 2.640 & 2.384\\
Mc-Ys & 2RR & 2.621 & 2.632 & 2.635 & 2.426\\
Mc-Ys & 3RR & 2.726 & 2.603 & 2.862 & 2.681\\
\hline
\end{tabular}
\begin{flushleft}
\small{{\bf Notes:} Station codes:
Mc -- Medicina, Tr -- Toru\'n, Wb -- Westerbork,\\Ys -- Yebes}
\end{flushleft}
\end{table}

In our proof-of-concept experiment we used an strontium optical lattice 
clock developed and operated at KL\,FAMO \citep{Morzynski2015} to synchronize
Toru\'n VLBI station during a test session as shown in Fig.\,7. The station was 
referenced to the clock over an optical frequency comb system and a 
15-km-long ELSTAB-based fibre link. The optical frequency comb was locked to 
the 698\,nm transition (429 228 066 418 008.3(2.1)\,Hz) from the optical 
clock. Both repetition rate and offset frequency of the comb were stabilized 
to the beat-note signal of one of the comb tooth and the so-called clock 
laser. The clock laser used in our optical atomic clock is stabilized to the 
ultra-stable cavity made of 100\,mm-long ultra-low expansion glass spacer 
and fused silica mirrors. The instability of our clock laser, when it is not 
referenced to the atomic transition, is below $2\times10^{-14}$ on time 
scales of 2 to $10^4$\,sec. The clock laser frequency is digitally locked to 
the frequency of strontium 698\,nm clock transition. The clock laser 
frequency stabilization is performed by a frequency shifter that also 
removes slow linear frequency drift of the ultra-stable cavity. The 
frequency corrections are derived from the excitation probability on both 
sides of the atomic line \citep{Morzynski2013}. The sample of atoms in the optical
lattice is 
produced and interrogated every 1.3\,sec., therefore the servo loop cycle in our 
set-up is equal to 2.6\,sec. The instability in terms of the Allan deviation of 
the locked clock laser starts to decrease after few seconds averaging with a 
slope close to $\uptau^{-1/2}$, where $\uptau$ is the averaging period.

With the frequency comb repetition rate actively stabilized to the optical 
atomic clock, a 10\,MHz signal can be produced by the Direct Digital 
Synthesizer (DDS) that is inside the comb stabilization loop (see Fig.\,7). 
Hence the 10\,MHz signal is locked to the frequency of strontium 698\,nm clock 
transition and its stability is not deteriorated by any other external RF 
sources. The 10\,MHz signal was subsequently transmitted to the Toru\'n VLBI 
station over a stabilized fibre link.

\begin{figure*}
\includegraphics[width=\linewidth]{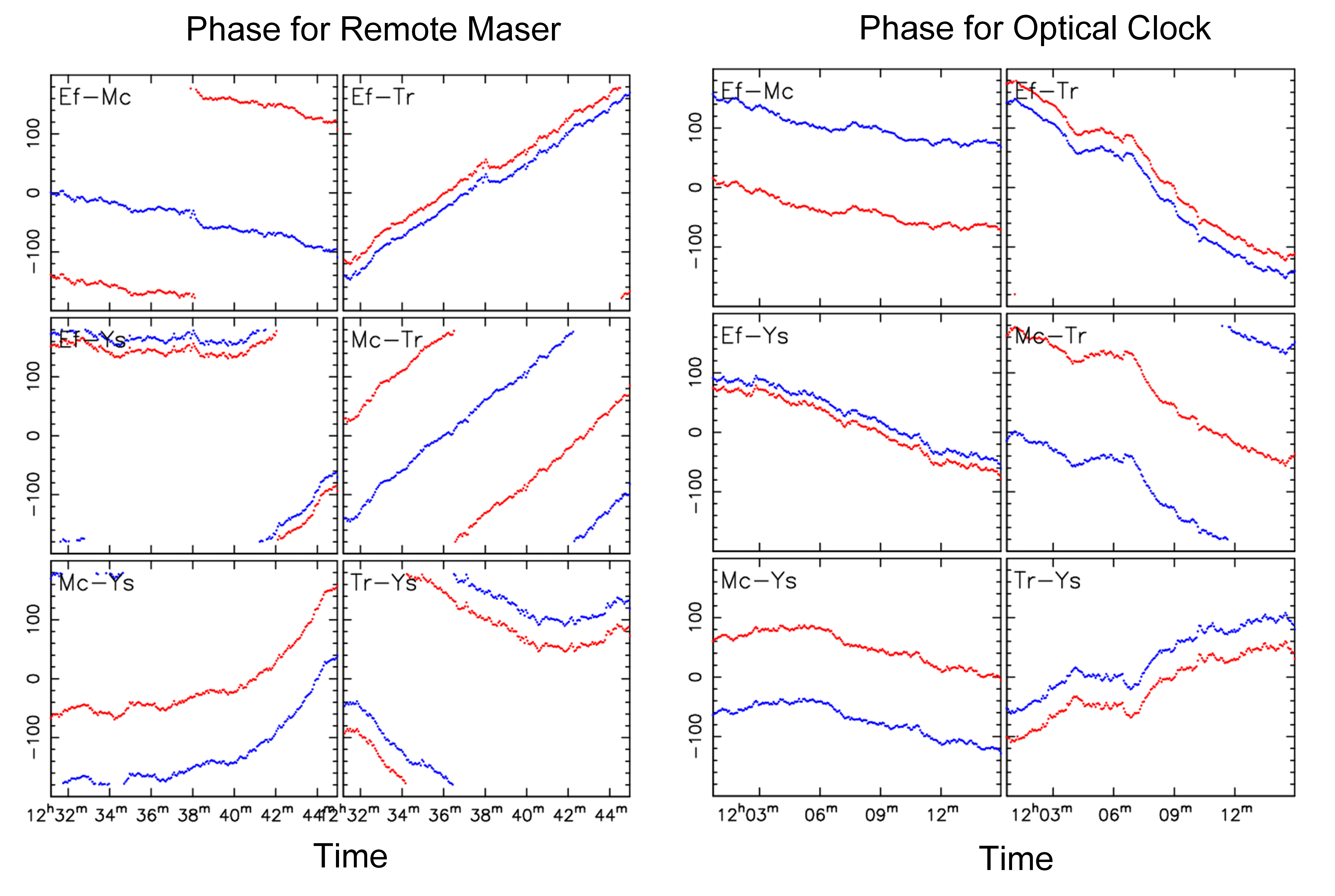}
\caption{Residual interferometer phase for two methods of providing T\&F 
control to the Toru\'n VLBI station during the experiment of 15~March~2016.
The $x$-axis is UTC and the $y$-axis is degrees of phase in 
both polarizations of the fourth sub-band (polarization colour-coding: red = 
RCP, blue = LCP), vector averaged over the middle 80\% of the sub-band. Each 
four-second integration interval is plotted as a separate point. The baselines 
are arranged such that those including Toru\'n are in the right-hand column 
for each of the T\&F control methods.}
\end{figure*}

We conducted a proof-of-concept experiment on 15~March~2016 during the
test time preceding a regular \hbox{e-VLBI} session of the European VLBI Network
(EVN). Besides the Toru\'n
VLBI station, the participating telescopes included Effelsberg (DE), Medicina 
(IT), and Yebes (ES). The observing format comprised eight dual-polarization 
16\,MHz sub-bands, together spanning 4926.49\,MHz to 5054.49\,MHz. The target 
throughout the observation was the bright radio quasar 0234+285. The 
tactics of this test observation were essentially the same as those for the 
test described in Sect.\,3. The two methods for T\&F control at Toru\'n VLBI 
station were the operational remote HM at AOS and the optical lattice clock 
described above. Once the clock-model coefficients were adjusted, we swapped
T\&F control to the optical lattice clock for 
three 15\,min scans and then swapped back to the remote HM. As in the 
previous test, in addition to the real-time \hbox{e-VLBI} correlation, the raw
data from the telescopes were also recorded at JIVE (two 15\,min scans for the 
remote HM).

Figure\,8 shows representative behaviour of the residual interferometer phases, 
plotting one 15\,min scan per domain for both hands of polarization for 
one sub-band. A feature that is typical of phase variations induced by 
propagation can be seen around 12:05--12:07 UTC for baselines to Toru\'n in the 
optical clock plots. At an observing frequency of 5\,GHz a cycle of phase 
corresponds to 200\,ps, thus the approximately 40-degree hiccup is equivalent 
to about 22\,ps, or to about 7\,mm of differential tropospheric path length.
We considered the same phase-difference statistics to characterize the phase 
noise achieved for each method of T\&F control at Toru\'n. Table\,2 shows the 
results for four of the 16 sub-band/polarizations for the case of the 
difference-pair statistic. Again, we have successfully proved that on these 
timescales the remote optical atomic clock can provide the operational 
synchronization of radio telescopes during VLBI observations.

\section{Summary}

We describe a fully operational solution of remote 
synchronization of the Toru\'n VLBI station with T\&F signals generated in a 
dedicated UTC laboratory and provided via a stabilized fibre link. We have
demonstrated that the remote synchronization does not deteriorate the noise 
characteristics of the observations and eliminates the issues related to 
local HM frequency offset and/or drift. In the particular situation of the 
Toru\'n VLBI station, the obvious benefit is the replacement of the local HM 
beginning to show signs of imminent failure. This is why the Toru\'n VLBI 
station has used the remote T\&F delivery in all VLBI observations in which 
it has participated since the beginning of 2016.

\begin{table}
\caption{Phase noise in picoseconds for different baselines and sub-bands
for remote HM and optical clock.}
\centering
\begin{tabular}{l c c c c c}
\hline
\hline
baseline & sub-     & \multicolumn{2}{c}{remote HM}
                    & \multicolumn{2}{c}{opt clock}\\
         & band     & noise & noise & noise & noise\\
         & and      & rms & rms & rms & rms \\
         & polarity & odd & even & odd & even \\
\hline
Ef-Tr & 0RR & 1.475 & 1.530 & 1.342 & 1.373\\
Ef-Tr & 1RR & 1.424 & 1.539 & 1.357 & 1.363\\
Ef-Tr & 2RR & 1.443 & 1.547 & 1.361 & 1.367\\
Ef-Tr & 3RR & 1.439 & 1.531 & 1.352 & 1.325\\
\hline
Mc-Tr & 0RR & 1.333 & 1.780 & 1.594 & 1.616\\
Mc-Tr & 1RR & 1.249 & 1.734 & 1.671 & 1.582\\
Mc-Tr & 2RR & 1.301 & 1.680 & 1.619 & 1.533\\
Mc-Tr & 3RR & 1.267 & 1.714 & 1.483 & 1.615\\
\hline
Tr-Ys & 0RR & 2.018 & 2.041 & 1.938 & 1.906\\
Tr-Ys & 1RR & 1.993 & 1.984 & 1.995 & 1.880\\
Tr-Ys & 2RR & 1.952 & 2.054 & 2.077 & 1.894\\
Tr-Ys & 3RR & 2.023 & 2.061 & 2.026 & 1.844\\
\hline
Ef-Mc & 0RR & 1.243 & 1.562 & 0.969 & 1.073\\
Ef-Mc & 1RR & 1.212 & 1.513 & 0.965 & 1.085\\
Ef-Mc & 2RR & 1.204 & 1.576 & 0.977 & 1.056\\
Ef-Mc & 3RR & 1.198 & 1.552 & 1.003 & 1.042\\
\hline
Ef-Ys & 0RR & 1.736 & 1.969 & 1.343 & 1.401\\
Ef-Ys & 1RR & 1.695 & 1.975 & 1.358 & 1.403\\
Ef-Ys & 2RR & 1.713 & 1.959 & 1.365 & 1.418\\
Ef-Ys & 3RR & 1.722 & 1.938 & 1.330 & 1.393\\
\hline
Mc-Ys & 0RR & 1.568 & 1.589 & 1.250 & 1.172\\
Mc-Ys & 1RR & 1.587 & 1.598 & 1.341 & 1.226\\
Mc-Ys & 2RR & 1.594 & 1.512 & 1.221 & 1.253\\
Mc-Ys & 3RR & 1.552 & 1.599 & 1.296 & 1.193\\
\hline
\end{tabular}
\begin{flushleft}
\small{{\bf Notes:} Station codes:
Ef -- Effelsberg, Mc -- Medicina, Tr -- Toru\'n,\\Ys -- Yebes}
\end{flushleft}
\end{table}

We also show the results of a proof-of-concept experiment with 
synchronization based on the strontium optical lattice clock, where the optical 
frequency was down-converted to the RF domain by an optical frequency comb 
and then delivered to the Toru\'n VLBI station via a stabilized fibre link. 
This experiment demonstrates the operational usefulness of the optical clock 
in VLBI observations and may be regarded as a first step towards more 
complex experiments that might show the noticeable benefits from using 
optical clocks instead of HMs, the traditional T\&F standards used in VLBI 
for decades. Our long-term strategy is to deliver an optical carrier instead 
of 10\,MHz signal to the Toru\'n VLBI station and to operate the optical 
frequency comb and down-conversion system locally at the station. In such 
circumstances, it would be possible to increase the resulting electrical 
reference frequency from 10\,MHz used in the proof-of-concept experiment to 
much higher value (of the order of several gigahertz). As an ultimate step, 
we plan to arrange the experimental session in cooperation with at least one 
other VLBI station also synchronized with an optical clock and to perform 
observation in the millimetre-wave range. These efforts combined would give 
final evidence that VLBI observations synchronized by optical clocks can 
outperform those carried out using HMs in terms of stability and accuracy.

We believe that the work presented here, together with other similar 
experiments described by \citet{Clivati2015}, \citet{Perini2016}, and
\citet{Clivati2017},
will start a discussion within the VLBI community on the potential benefits
of a traceable, remote synchronization of VLBI stations either using HMs
or even more stable and accurate optical clocks.

\begin{acknowledgements}

We thank the teams of the Westerbork, Medicina, and Yebes VLBI stations 
for their willingness to participate in the test of 18~December~2015 
organized in the form of a dedicated VLBI session outside the regular EVN 
time. This work was supported by Polish National Science Center (NCN) under
projects no. 2014/15/B/ST7/00471 and 2015/17/B/ST7/03628. P. Morzy\'nski
was partly supported by Polish National Science Center (NCN) under project
no. 2014/15/D/ST2/05281. Support has been received from the project EMPIR
15SIB03 OC18. This project has received funding from the EMPIR programme
co-financed by the Participating States and from the European Union’s Horizon
2020 research and innovation programme. This research is a part of the
programme of the National Laboratory FAMO in Toru\'n, Poland.

\end{acknowledgements}

{}

\end{document}